\documentclass[12pt]{article}
\usepackage{setspace}
\usepackage{epsfig}
\usepackage{amssymb}
\def\be{\begin{equation}}
\def\ee{\end{equation}}
\def\ba{\begin{array}}
\def\ea{\end{array}}
\def\beqn{\begin{eqnarray}}
\def\eeqn{\end{eqnarray}}

\def\bt{\begin{tabular}}
\def\et{\end{tabular}}
\def\bc{\begin{center}}
\def\ec{\end{center}}
\begin{document}
\title{Implications of unitarity and precision measurements on CKM matrix elements}
\author{Gulsheen Ahuja$^1$, Manmohan Gupta$^1$, Sanjeev Kumar$^2$,
 Monika Randhawa$^3$ \\
\\{$^1$ \it Department of Physics, Centre of Advanced Study, P.U.,
Chandigarh, India.}\\{$^2$ \it Department of Physics, H.P.U.,
Shimla, India.}\\{$^3$ \it University Institute of Engineering and
Technology, P.U., Chandigarh, India.}\\{\it Email:
mmgupta@pu.ac.in}}
 \maketitle
\begin{abstract}
Unitarity along with precision measurements of sin$\,2\beta$,
$V_{us}$ and $V_{cb}$ allows one to find a lower bound $V_{ub}\geq
0.0035$ which, on using the recently measured angle $\alpha$ of
the unitarity triangle, translates to $V_{ub}= 0.0035\pm 0.0002$.
This precise value, stable for a good deal of changes in $\alpha$,
along with CP violating phase $\delta$ found from unitarity allows
the construction of a `precise' CKM matrix. The above unitarity
based value of $V_{ub}$ is in agreement with the latest exclusive
value used as input by UTfit, CKMfitter, HFAG, however underlines
the so called `tension' faced by the latest inclusive
$V_{ub}=0.00449 \pm 0.00033$. Further, using this inclusive value
of $V_{ub}$ along with the latest sin$\,2\beta$, one finds
$\delta=23 ^{\rm o}- 39 ^{\rm o}$, again in conflict with $\delta$
measured in B-decays. The calculated ranges of the elements of the
CKM matrix are in excellent agreement with those obtained recently
by UTfit, CKMfitter and HFAG. Also, the ratio
$\frac{V_{ts}}{V_{td}}$ is in agreement with its latest measured
value, whereas there is some disagreement between the `measured'
and the calculated $V_{td}$ values.
 \end{abstract}

In the last few years, extremely important developments have taken
place in the context of phenomenology of Cabibbo-Kobayashi-Maskawa
(CKM) matrix \cite{ckm}, both from theoretical as well as
experimental point of view. The precise measurement of CP
violating parameter sin$\,2\beta$ \cite{s2b}-\cite{hfag} and a
fairly precise measurement of angle $\alpha$ \cite{gronau} of the
unitarity triangle in B-decays have allowed a precise
determination of the phase of the CKM matrix. Several detailed and
extensive phenomenological analyses
\cite{utfit}-\cite{hfag},\cite{pdgnew} have allowed us to conclude
that the single CKM phase looks to be a viable solution of CP
violation not only in the case of K-decays but also in the context
of B-decays, at least to the leading order. On the one hand, this
situation looks highly satisfactory from the Standard Model (SM)
point of view, on the other hand, it has also triggered intense
amount of activity on the theoretical as well as experimental
front for finding clues to New Physics (NP).

Several authors \cite{bnp}-\cite{ligeti} have suggested possible
strategies for deciphering NP in the context of CKM phenomenology.
One possible way to observe NP is by discovering violations of
unitarity, as emphasized by Buras \cite{bnp}. In this context, it
needs to be noted that the persistent $2\sigma$ violation of
unitarity by the first row elements of the CKM matrix has been
eliminated by improving the precision in the measurement of
$V_{us}$ \cite{utfit,vus}. This, however, has also triggered a
great deal of interest in measuring the other CKM elements to
better and better accuracy for testing unitarity of the CKM
matrix. There are several CKM elements and phenomenological
parameters, e.g., $V_{us}, V_{cb}, V_{ud}, \alpha, \beta$, etc.,
wherein the error bars are limited to only a few percent, however,
this is not true in the case of several other CKM elements such as
$V_{ub},V_{cs}, V_{ts}$ and $V_{td}$. A precise knowledge of these
elements would not only test unitarity to better and better level,
but would also provide clues to the possibility of existence of
NP. It may be noted that the elements $V_{ts}$ and $V_{td}$, under
the present circumstances, can only be measured indirectly,
whereas the elements $V_{ub}$ and $V_{cs}$ can be measured through
tree level decays, therefore, one would expect considerable
improvements in these in the near future. At present, even after
recent updating by various groups, the situation regarding
$V_{ub}$ remains largely unclear. As per PDG 2006 \cite{pdgnew}
the exclusive and inclusive values respectively are $V_{ub} =
0.00384 + 0.00067 - 0.00049$ and $V_{ub} = 0.00440 \pm 0.00029 \pm
0.00027$, whereas the October 2006 (including Summer (ICHEP06)
updates) updated analysis by UTfit Collaboration \cite{utfit},
also agreed by CKM05 Workshops \cite{cfitter} and HFAG\cite{hfag},
uses as inputs exclusive $V_{ub}=0.0035 \pm 0.0004$ and inclusive
$V_{ub}=0.00449 \pm 0.00033$. Keeping in mind the significance of
the difference between the exclusive and inclusive values of
$V_{ub}$, the UTfit \cite{utfit} carries out a separate analyses
for these. This point of view has also been advocated by several
authors \cite{ali}.

The unprecedented accuracy in the measurement of sin$\,2\beta$
\cite{s2b}-\cite{hfag}, $V_{us}$ \cite{utfit,vus} and $V_{cb}$
\cite{utfit} provides a strong motivation for carrying out a fine
grained analysis for testing CKM paradigm to better and better
accuracy as well as in search for clues to situations which have
potential seeds for NP. Similarly, a very recent precise
measurement of $\Delta M_{B_s}$ \cite{mbs} would not only have
implications for CKM elements $V_{ts}$ and $V_{td}$ but would also
have implications for other CKM phenomenological parameters
\cite{utfit}. In this context, unlike the several global analyses
\cite{utfit}-\cite{hfag},\cite{pdgnew} carried out recently, it
would perhaps be desirable to fine tune the implications of each
of the vital inputs of CKM paradigm separately along with the
precision measurements on the CKM matrix elements and other
phenomenological parameters. To this end, an analysis emphasizing
unitarity of the CKM matrix and the precisely measured CKM
parameters as well as some of the over constraining measurements
would be very desirable.

The purpose of the present communication is to study the
implications of unitarity along with the well measured $V_{us}$,
$V_{cb}$, sin$\,2\beta$, and angle $\alpha$ of the unitarity
triangle on some of the lesser known elements of the CKM matrix
such as $V_{ub}, V_{cs}, V_{ts}$ and $V_{td}$. In particular, one
would like to examine in detail the implications of unitarity
along with recently refined sin$\,2\beta$ on exclusive and
inclusive values of $V_{ub}$ and CP violating phase $\delta$.
Using minimal inputs, e.g., unitarity and other well measured
quantities, it would also be of interest to explore the
possibility of constructing a `precise' CKM matrix. Apart from
examining the compatibility of over constraining measurements, we
would also like to find the unitarity based predictions for
Jarlskog's rephasing invariant parameter $J$ as well as the
Wolfenstein-Buras parameters $\overline{\rho}$ and
$\overline{\eta}$.

Most of the present day analyses, related to CKM phenomenology,
have been carried out using the Wolfenstein-Buras parametrization
\cite{wolfbur} of the CKM matrix. However, in the present case, as
the emphasis is on unitarity therefore we find it more convenient
to use the PDG representation of the CKM matrix, wherein the
 unitarity is built-in. For ready reference as well as to facilitate discussion of
 results, we begin by considering the quark mixing matrix,
\be  V_{{\rm CKM}}
  = \left( \ba{ccc} V_{ud} &  V_{us} & V_{ub} \\  V_{cd} &  V_{cs} &
   V_{cb} \\  V_{td} &  V_{ts} &  V_{tb} \ea \right),  \label{ckm}  \ee
which in the PDG representation, involving angles $\theta_{12},
\theta_{23}, \theta_{13}$ and phase $\delta$ \cite{pdgnew} is
given as
  \be V_{{\rm CKM}}=\left( \ba{ccl} c_{12} c_{13} & s_{12} c_{13} &
  s_{13}e^{-i \delta} \\ - s_{12} c_{23} - c_{12} s_{23}
  s_{13} e^{i \delta_{13}} & c_{12} c_{23} - s_{12} s_{23}
  s_{13} e^{i \delta_{13}} & s_{23} c_{13} \\ s_{12} s_{23} - c_{12}
  c_{23} s_{13} e^{i \delta_{13}} & - c_{12} s_{23} - s_{12} c_{23}
  s_{13} e^{i \delta_{13}} & c_{23} c_{13} \ea \right), \label{mm} \ee
with $c_{ij}={\rm cos}\, \theta_{ij}$ and $s_{ij}={\rm sin}\,
\theta_{ij}$, for $i,\,j=1,2,3$. In this representation, one can
consider up to 4th decimal place $V_{us} = s_{12}$ and $V_{cb} =
s_{23}$, whereas $|V_{ub}| = s_{13}$, henceforth $|V_{ub}|$ would
be written as $V_{ub}$.

Unitarity of the $V_{{\rm CKM}}$ implies nine relations, three in
terms of normalization conditions also referred to as `weak
unitarity conditions', and the other six are usually expressed
through unitarity triangles in the complex plane. Because of the
strong hierarchical nature of the CKM matrix elements as well as
the limitations imposed by the present level of measurements, it
is difficult to study the implications of normalization relations,
therefore, the six non-diagonal relations are used to study the
implications of unitarity on CKM phenomenology. Out of the six,
four triangles implied by these relations are highly skewed and it
is difficult to study their implications \cite{mon,botella} with
the present knowledge of the CKM matrix elements. The implications
of the other two are usually studied through the triangle
expressed by the relation
  \be V_{ud} V_{ub}^* + V_{cd} V_{cb}^* + V_{td}
V_{tb}^* =0\,,\label{db} \ee also referred to as $db$ triangle.
The angles of this triangle, in terms of $V_{{\rm CKM}}$ elements,
mixing angles and CP violating phase $\delta$ \cite{pdgnew},
related to CP asymmetries, are expressed as
 \begin{eqnarray}\alpha\equiv{\rm arg}\left[-\frac{V_{td} V_{tb}^*}{V_{ud}
 V_{ub}^*}\right]=\tan^{-1}\left[\frac{s_{12}
 s_{23}\, {\rm sin}\, \delta}{c_{12} c_{23} s_{13}-s_{12} s_{23}\, {\rm cos}\,\delta}\right]
  ,\label{alpha} \end{eqnarray}
 \begin{eqnarray} \beta\equiv{\rm arg}\left[-\frac{V_{cd} V_{cb}^*}{V_{td}
 V_{tb}^*}\right]=\tan^{-1}\left[\frac{c_{12}
 s_{12} s_{13}\, {\rm sin}\,\delta}{c_{23} s_{23} (s_{12}^2-c_{12}^2 s_{13}^2)-c_{12} s_{12} s_{13}
 (c_{23}^2-s_{23}^2)\, {\rm cos}\,\delta}\right]
,\label{beta} \end{eqnarray}
 \begin{eqnarray} \gamma\equiv{\rm arg}\left[-\frac{V_{ud} V_{ub}^*}{V_{cd}
 V_{cb}^*}\right]=\tan^{-1}\left[\frac{s_{12}
 c_{23}\, {\rm sin}\,\delta}{c_{12} s_{23} s_{13}+s_{12} c_{23}\, {\rm cos}\,\delta}\right]
\label{gamma}. \end{eqnarray} To obtain information about the CP
violating phase $\delta$ from the experimentally well determined
angle $\beta$ one can express equation (\ref{beta}) as
\be
{\rm tan}\,\frac{\delta}{2} = \frac{A - \sqrt{A^2-(B^2-A^2
C^2){\rm tan}^2 \beta}}{(B+AC){\rm tan}\,\beta}, \label{tand} \ee
where $A=c_{12} s_{12} s_{13}$, $B=c_{23} s_{23}(s_{12}^2-c_{12}^2
s_{13}^2)$ and $C=c_{23}^2-s_{23}^2$. Using $s_{12}^2 \gg c_{12}^2
s_{13}^2$ and $s_{23}^2 \ll
 c_{23}^2$, the above relation can be re-expressed as
\be
\delta~=-\beta+{\rm
sin}^{-1}\left(\frac{s_{12}s_{23}}{c_{12}s_{13}}{\rm
sin}\beta\right), \label{delb}\ee which can also be written as
\be
\frac{{\rm sin}\,(\delta+\beta)}{{\rm
sin}\,\beta}=\frac{s_{12}s_{23}}{c_{12}s_{13}}. \label{dpb}\ee
From equation (\ref{gamma}), one can easily show that $\gamma =
\delta$ with an error of around 2 $\%$, therefore, using the
closure property of the angles of the triangle,
$\alpha+\beta+\gamma=\pi$, the above
 equation can be written as
\be
s_{13}=\frac{s_{12}s_{23}\,{\rm sin}\,\beta}{c_{12}\,{\rm
sin}\,\alpha}, \label{s13ab} \ee which can also be derived from
equation (\ref{alpha}) by using the closure property of the
triangle. Equation (\ref{dpb}) can be used to provide a lower
bound on $s_{13}$, e.g.,
\be
s_{13}~\geq~\frac{s_{12}s_{23}}{c_{12}}\,{\rm sin}\,\beta.
\label{lbs13} \ee

 \begin{table}
\begin{tabular}{|l|l|l|}  \hline
Parameter & Latest (October 2006) Values & PDG 2006 values
\cite{pdgnew}
\\ \hline$V_{us}$ & 0.2258 $\pm$ 0.0014  & 0.2257 $\pm$ 0.0021\\
$V_{cb}$ & 0.0416 $\pm$ 0.0007  & 0.0416 $\pm$ 0.0006 \\
 ${\rm sin}\,2\beta$ & 0.675 $\pm$ 0.026
& 0.687 $\pm$ 0.032\\
 $\beta$ & (21.24 $\pm$ 1.01)$^{\rm
o}$& (21.7 $\pm$ 1.2)$^{\rm o}$\\
 $\alpha$ & (91.0 $\pm$ 7.0 $\pm$ 3.0)$^{\rm o}$&
 (99.0 + 13.0 - 8.0)$^{\rm o}$\\
 $\gamma$ or $\delta$ & $(63.0 + 15.0 - 12.0)^{\rm o}$&
 $(63.0 + 15.0 - 12.0)^{\rm o}$\\
 $V_{ub}$~(excl.) & 0.0035 $\pm$ 0.0004& 0.00384 + 0.00067 - 0.00049\\
 $V_{ub}$~(incl.) & 0.00449 $\pm$ 0.00033 & 0.00440 $\pm$
 0.00029 $\pm$ 0.00027\\
 \hline
 \end{tabular}
\caption{The PDG 2006 \cite{pdgnew} measured values and the latest
October 2006 (including Summer (ICHEP06) updates) input values
used by UTfit Collaboration \cite{utfit}, also agreed by CKM05
Workshops \cite{cfitter} and HFAG\cite{hfag}. The latest values of
$\alpha$ and $\gamma$ are from \cite{gronau} and
\cite{cfitter,glw} respectively.}
 \label{input}
 \end{table}

Before we discuss the details of our analysis, in table
\ref{input} we present the PDG 2006 \cite{pdgnew} measured values
and the latest input values of some of the CKM elements and the
angles of the unitarity triangle used by UTfit Collaboration
\cite{utfit}, also agreed by CKM05 Workshops \cite{cfitter} and
HFAG\cite{hfag}. The values of angles $\alpha$ and $\gamma$ have
not been used as inputs by UTfit Collaboration, therefore we use
their latest values from \cite{gronau} and \cite{cfitter,glw}
respectively.

\begin{figure}[hbt]
\centerline{\psfig{figure=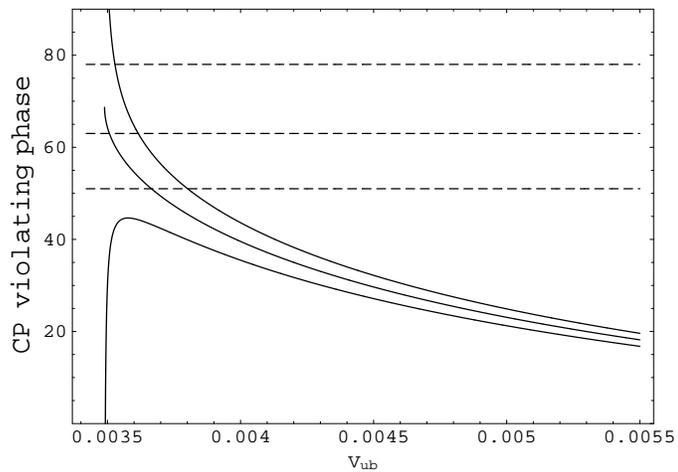,width=3.5in,height=2.5in}}
\caption{Plot showing variation of $V_{ub}$ versus CP violating
phase $\delta$, obtained by using equation (\ref{tand}). The
central solid line corresponds to mean value of input parameters,
whereas the other 2 lines correspond to 1$\sigma$ variations. The
horizontal dashed lines correspond to $\delta=(63.0 + 15.0
-12.0)^{\rm o}$, the central line corresponds to mean value.}
 \label{vubdel}
  \end{figure}

To begin with, we study the implications of unitarity and
precisely measured recently improved sin$\,2\beta$ on CP violating
phase $\delta$ and $V_{ub}$. On examining unitarity based equation
(\ref{tand}), we find that $\delta$ is dependent on $V_{us}$,
$V_{cb}$, angle $\beta$ as well as it involves $V_{ub}$. Using
this equation, in figure \ref{vubdel} we have plotted the CP
violating phase $\delta$ versus $V_{ub}$, also included in the
figure is the experimentally measured $\delta=(63.0 + 15.0
-12.0)^{\rm o}$ shown by horizontal dashed lines, inclusive of
results of various global analyses. The solid central line depicts
$\delta$ obtained by using the mean values of $V_{us}$, $V_{cb}$
and  sin$\,2\beta$ whereas the outer lines correspond to the
1$\sigma$ ranges of these inputs. A general look at the figure
reveals several interesting points, e.g., for values of $V_{ub} >
0.00355$, the central value of $\delta$ shows a smooth decline as
well as the range of $\delta$ gets narrower and narrower with
increasing $V_{ub}$, however for $V_{ub} < 0.00355$ it seems that
there is a sharp broadening of the $\delta$ range, with no
restriction on $\delta$ when $V_{ub} < 0.0035$. From the graph one
finds that the 1$\sigma$ range of the recent inclusive value of
$V_{ub}$, as given in table \ref{input} restricts $\delta$ to $23
^{\rm o}- 39 ^{\rm o}$, whereas the mean value of the recent
exclusive value does not constrain $\delta$, however the upper
limit of the 1$\sigma$ range of the exclusive value provides only
a lower bound $\delta > 38 ^{\rm o}$. In conclusion, we would like
to emphasize that the precisely known sin$\,2\beta$, for the
inclusive value of $V_{ub}$ implies a narrow range for $\delta$,
whereas for the exclusive value of $V_{ub}$ it implies only a
lower bound on $\delta$. Figure \ref{vubdel} can also be used for
constraining $V_{ub}$ for particular values of $\delta$, e.g., the
range given in the table implies $V_{ub} < 0.0038$. One may wonder
whether a similar analysis can be carried out using the latest
measured value of angle $\alpha$. We have carried out such an
analysis, however it does not lead to any new conclusions.

Our conclusions about $V_{ub}$ and $\delta$ can be sharpened
further by using other unitarity based relations. To this end,
equation (\ref{dpb}) allows $V_{ub}$ to be expressed in terms of
the well determined quantities $V_{us}$, $V_{cb}$ and sin$\beta$.
Interestingly, in case $\delta$ is also a well measured quantity,
then this equation immediately leads to a precise prediction for
$V_{ub}$. However, even in the case where $\delta$ is not well
determined, one can use equation (\ref{lbs13}) to obtain a
rigorous lower bound on $V_{ub}$. A simple calculation using the
mean values of input parameters immediately leads one to
\be
V_{ub}\geq 0.0035. \label{alb} \ee It may be noted that this bound
is independent of the value of $\delta$ as well as contamination
of NP in the measurement of $\delta$. Interestingly, equation
(\ref{dpb}) can also be used to show $V_{ub}\leq 0.00402$, found
by using the lower limits of $\delta$ and $\beta$ as given in the
table.

Our predictions regarding $V_{ub}$ can be refined further in case
we incorporate angle $\alpha$ of the unitarity triangle, measured
from $B \rightarrow \pi \pi$ and $B \rightarrow \rho \rho$ decays.
Using its present consensus value \cite{gronau}, as given in table
\ref{input}, from equation (\ref{s13ab}) one finds \be
 V_{ub}= 0.0035\pm 0.0002. \label{s13r} \ee
Interestingly, this precise value is in full agreement with the
recently used input value of exclusive $V_{ub}$ by UTfit
\cite{utfit}, however it has much smaller error bars. The above
value of $V_{ub}$ is a consequence of unitarity and the precisely
measured elements $V_{us}$, $V_{cb}$ and angles $\beta$ and
$\alpha$. It may also be emphasized that this value is quite
insensitive to a change in the value of angle $\alpha$. In fact,
even if the mean value of $\alpha$ changes by more than 20 $\%$,
still $V_{ub}$ would register a variation of only a few percent.
Also, refinements in the measurement of $\delta$ would not affect
the value of $V_{ub}$ in equation (\ref{s13r}) as $\delta$ along
with sin$\,2\beta$ gives only a lower bound on $V_{ub}$, mentioned
in equation (\ref{alb}). Therefore, the above prediction of
$V_{ub}$ can be considered as a rigorous and robust consequence of
unitarity.

The above discussion also underlines the fact that precisely
measured sin$\,2\beta$ along with $V_{us}$ and $V_{cb}$ does not
lead to any well defined conclusion regarding $\delta$ because of
the persistent difference between exclusive and inclusive values
of $V_{ub}$. Therefore, to find unitarity based $\delta$ one has
to use the closure property of the angles of the unitarity
triangle. Using the well measured angles $\alpha$ and $\beta$, one
obtains
 \be \delta=67.8^{\rm o}\pm 7.3^{\rm o}. \label{delta}\ee
This unitarity based value of $\delta$ is compatible with the
directly measured value in $B^\pm \rightarrow D K^\pm$ decays
\cite{glw} as well as with the recently obtained $\delta$
\cite{buras1} from the $B \rightarrow \pi \pi$ and $B \rightarrow
\pi K$ decays. It may also be mentioned that this value is
compatible with the $\delta$ bound given by exclusive $V_{ub}$, as
obtained from figure \ref{vubdel}, however does not agree with the
$\delta$ range obtained for inclusive $V_{ub}$.

After having found $V_{ub}$ and $\delta$ from unitarity, one would
like to construct the entire CKM matrix which is obtained at
1$\sigma$ C.L. as follows
 \be V_{{\rm CKM}} = \left( \ba{ccc}
  0.9738 ~{\rm to}~ 0.9745 &   0.2244~ {\rm to} ~0.2272 &  0.0033 ~{\rm to} ~0.0036 \\
 0.2243 ~{\rm to}~ 0.2270  &   0.9730 ~{\rm to}~ 0.9736    &  0.0409~ {\rm to}~ 0.0423\\
0.0082 ~{\rm to}~ 0.0091  &  0.0401 ~{\rm to}~ 0.0415 &  0.9990~
{\rm to}~ 0.9991 \ea \right). \label{1sm} \ee

It may be mentioned that this matrix is free from contamination by
NP to the extent that the measured values of angles $\alpha$ and
$\beta$ are free from NP effects. Also, it needs to be emphasized
that this has been constructed by using minimal inputs such as
$V_{us}$, $V_{cb}$, $V_{ub}$, sin$\,2\beta$ and the unitarity
based PDG parametrization, however without incorporating the full
constraints due to unitarity. A general look at the matrix reveals
that the ranges of CKM elements obtained here are quite compatible
with those obtained by recent global analyses. In particular, the
ranges found here are in excellent agreement with those emerging
from global fits by UTfit, CKMfitter and HFAG. This perhaps
indicates that unitarity plays a key role even in the case of
global analyses. However, it must be mentioned that although the
matrix presented here agrees well with the one given by PDG, yet
there is a slight disagreement in the case of $V_{ub}$ and
$V_{td}$. The discrepancy in $V_{ub}$ can be easily understood as
the $V_{ub}$ value used here is somewhat lower than the average
$V_{ub}$ value considered by PDG 2006. The disagreement in the
value of the element $V_{td}$, sensitive to both loop and NP
effects, suggests the need for further experimental scrutiny in
this case. An experimental confirmation of the values of the CKM
elements would strengthen the present unitarity based analysis as
well as its predictions regarding $V_{ub}$ and $\delta$.

For the sake of completeness and better appreciation of the
present results, we have evaluated the Jarlskog's rephasing
invariant parameter $J$ and the Wolfenstein-Buras parameters
$\overline{\rho}$ and $\overline{\eta}$ by expressing these in
terms of the mixing angles and the CP violating phase $\delta$.
Using the experimental values of $V_{us}$, $V_{cb}$ as well as the
unitarity based values of $V_{ub}$ and $\delta$ found above, we
obtain \be J=(2.95 \pm 0.22) 10^{-5} \label{j}, \ee
 \be \overline{\rho}=0.14 \pm 0.04~~~~~~~~~~{\rm
and}~~~~~~~~~~~ \overline{\eta}=0.34 \pm 0.02 \label{rhoeta}. \ee
Interestingly, the value of $\overline{\eta}$ is in complete
agreement with those found by recent global analyses
\cite{utfit}-\cite{hfag},\cite{pdgnew}, whereas in case of
$\overline{\rho}$ the value found here agrees with the one
obtained by UTfit Collaboration \cite{utfit}.

The present analysis brings out several points which need to be
emphasized. The so called `tension' between the precisely known
sin$\,2\beta$ and the inclusive value of $V_{ub}$, as has already
been observed by several authors \cite{utfit,ligeti,tension},
becomes quite evident in the present analysis. In the present
context, this tension gets depicted in the form of disagreement
between the value of $\delta$ implied  by inclusive $V_{ub}$ and
the measured value of $\delta$. From figure \ref{vubdel}, one
immediately finds that the $\delta$ value corresponding to
inclusive $V_{ub}$ comes out to be much smaller than the
experimentally measured $\delta$. This `tension' is also visible
in the form that the present unitarity based $V_{ub}$ is much
smaller than the inclusive value of $V_{ub}$, however is in
excellent agreement with the latest exclusive $V_{ub}$. Therefore,
the so called `tension' can also be seen as a disagreement between
the unitarity based/exclusive and inclusive values of $V_{ub}$. It
also becomes clear that in the case of PDG 2006, the exclusive
value used by them is somewhat higher than the one used by other
global analyses, therefore, they find corresponding reduction in
the so called `tension'.

The CKM matrix constructed above allows us to calculate the ratio
$\frac{V_{ts}}{V_{td}}$ which is expected to be free from hadronic
uncertainties. The present calculated value $4.69 \pm 0.23$ looks
to be quite precise and has an excellent overlap with $4.7 \pm
0.4$ \cite{klein}, found recently from precision measurements of
$\Delta M_{B_s}$. Also, the measured value of the ratio
$\frac{V_{ts}}{V_{td}}$ can be considered as an over constraining
check on the above unitarity based predictions. Interestingly,
this measurement also provides an indirect check on the unitarity
based $\delta$ value used here which can be seen as follows. One
can easily check that $V_{ts}$ is essentially independent of
$\delta$ and $V_{ub}$, therefore can be predicted quite accurately
from unitarity. The element $V_{td}$ has hardly any dependence on
$V_{ub}$ while it is known to be very much $\delta$ dependent,
therefore the ratio $\frac{V_{ts}}{V_{td}}$ measurement can be
considered to imply a precise value of $\delta$ which would fully
agree with the value considered here. This possibility also
ensures the validity of our $\delta$ dependent construction of CKM
matrix even if the error bars in $\alpha$ become larger leading to
larger error in the $\delta$ value found from the closure
relationship.

One would also like to emphasize that the present $V_{td}=0.0087
\pm 0.0004$ looks to be at variance with $V_{td}=0.0072 \pm 0.0008
$ \cite{ulrich} found from $\Delta {M_B}_d$. In case one takes the
present values of hadronic factors used in the calculation of
$\Delta {M_B}_d$ seriously, then this difference may indicate the
presence of NP in $B^0 - \overline{B}^0$ mixing. The above
mentioned conflict in the $V_{td}$ values gets further sharpened
in case we consider the recently obtained $\delta=74^{\rm o} \pm
6^{\rm o}$ \cite{buras1} from the $B \rightarrow \pi \pi$ and $B
\rightarrow \pi K$ decays. While this value would be compatible
with the implied $\delta$ bound found from the present unitarity
based $V_{ub}$, however would be in conflict with the one found
using inclusive $V_{ub}$, as well as would imply $V_{td}\sim
0.0091$, aggravating the above conflict further.

It also needs to be mentioned that a further precision in the
measurement of sin$\,2\beta$, needless to say, would have far
reaching implications for CKM phenomenology, particularly for CP
violating phase $\delta$ and $V_{ub}$. It should also be noted
that in case the value of $V_{ub}$ is found to be $\sim 0.0035$
then it will need a careful scrutiny for studying its implications
as around this value the behaviour of $\delta$ and $V_{ub}$ in
figure \ref{vubdel} depicts sharp changes.

A summary of our principal conclusions is as follows. Unitarity
along with precisely measured $V_{us}$, $V_{cb}$ and sin$\,2\beta$
leads to $V_{ub}\geq 0.0035$. In case one uses the measured value
of the angle $\alpha$ of the unitarity triangle, one finds
$V_{ub}= 0.0035 \pm 0.0002$, this precise value can be considered
as a rigorous prediction of unitarity along with the other
precisely measured quantities as it is almost independent of good
deal of changes in $\alpha$. This is in agreement with the latest
exclusive $V_{ub}=0.0035 \pm 0.0004$ used as input by UTfit,
CKMfitter and HFAG, however is in conflict with the latest
inclusive $V_{ub}=0.00449 \pm 0.00033$, bringing out the so called
`tension' faced by inclusive $V_{ub}$. Further, when this
inclusive $V_{ub}$ is used along with sin$\,2\beta$ one finds
$\delta=23 ^{\rm o}- 39 ^{\rm o}$, again in conflict with the
$\gamma$ or $\delta$ measured in B-decays.

Using unitarity based closure property of the angles of the
unitarity triangle, one can find almost precise $\delta$ which
along with other precisely known elements allows one to construct
an almost `precise' CKM matrix. The ranges of CKM elements of the
present matrix, constructed by using `minimal inputs', are in
excellent agreement with those emerging from global fits by UTfit,
CKMfitter and HFAG. Also, the ratio $\frac{V_{ts}}{V_{td}}$
\cite{klein} is in full agreement with its latest measured value,
whereas in the case of $V_{td}$ there is some disagreement with
its recent measured value \cite{ulrich}, perhaps indicating the
presence of NP. The unitarity based values of $J,~\overline{\rho}$
and $\overline{\eta}$ found here are in agreement with those found
by some latest global analyses.

\vskip 0.2cm {\bf Acknowledgements} \\ The authors would like to
thank A. Ali and I. Bigi for useful suggestions. M.G. and G.A.
would like to thank DAE, BRNS (grant No.2005/37/4/BRNS), India,
for financial support. S.K. acknowledges the financial support
provided by CSIR, India. M.R. would like to thank the Director,
UIET for providing facilities to work.

\vskip 5cm


\begin{thebibliography}{99}

\bibitem{ckm} N. Cabibbo, Phys. Rev. Lett. 10 (1963) 531; M.
Kobayashi, T. Maskawa, Prog. Theor. Phys. 49 (1973) 652.

\bibitem{s2b}B. Aubert {\it et al.}, [BABAR Collaboration], hep-ex/0607107,
K.-F. Chen {\it et al.}, [Belle Collaboration], hep-ex/0608039.

\bibitem{utfit}M. Bona {\it et al.}, [UTfit Collaboration], JHEP 0610 (2006)
081, updated results available at http://www.utfit.org/.

\bibitem{cfitter}J. Charles {\it et al.}, [CKMfitter Group], Eur.
Phys. J. C41 (2005) 1, updated results available at
http://ckmfitter.in2p3.fr/.

\bibitem{hfag}E. Barberio {\it et al.}, Heavy Flavor Averaging Group (HFAG),
hep-ex/0603003, updated results available at
http://www.slac.stanford.edu/xorg/hfag/.

\bibitem{gronau}M. Gronau, hep-ph/0510153, hep-ph/0607282.

\bibitem{pdgnew}W.-M. Yao {\it et al.}, J. Phys. G 33 (2006) 1,
updated results available at http://pdg.lbl.gov/.

\bibitem{bnp}A. J. Buras, hep-ph/0505175.

\bibitem{nir}Y. Nir, hep-ph/0510413.

\bibitem{gro}M. Gronau, Phys. Lett. B627 (2005) 82.

\bibitem{ligeti}Z. Ligeti, PoS LAT2005 (2005) 012.

\bibitem{vus}A. Czarnecki, W. J. Marciano, A. Sirlin, Phys. Rev. D70 (2004) 093006.

\bibitem{ali}Private communication from A. Ali and I. Bigi.

\bibitem{mbs}http://www-cdf.fnal.gov/physics/new/bottom/060406.blessed-Bsmix/BsMixingMeasurement.pdf.

\bibitem{wolfbur}L. Wolfenstein, Phys. Rev. Lett. 51 (1983) 1945;
A. J. Buras, M. E. Lautenbacher, G. Ostermaier, Phys. Rev. 50
(1994) 3433.

\bibitem{mon}M. Randhawa, V. Bhatnagar, P. S. Gill, M. Gupta, Mod.
Phys. Lett. A (2000) 2363; M. Randhawa, M. Gupta, Phys. Lett. B516
(2001) 446.

\bibitem{botella}F. J. Botella, G. C. Branco, M. Nebot, M. N. Rebelo,
Nucl. Phys. B725 (2005) 155.

\bibitem{glw}A. Hocker {\it et al.}, Eur. Phys. J. C21
(2001) 225, updated results available at
http://ckmfitter.in2p3.fr/.

\bibitem{buras1}A. J. Buras, R. Fleischer, S. Recksiegel,
 F. Schwab, Eur. Phys. J. C45 (2006) 701.

\bibitem{tension}G. Paz, hep-ph/0612077, updated results available
at http://www.slac.stanford.
edu/xorg/ckmfitter /plots
beauty06/ckmEval results beauty06 .ps.gz.

\bibitem{klein}K. Kleinknecht, B. Renk, Phys. Lett. B639 (2006) 612.

\bibitem{ulrich}U. Nierste, Int. J. Mod. Phys. A21 (2006) 1724.

\end{thebibliography}
\end{document}